# Magnetic Field Dependent Behavior of the CDW ground state in $Per_2M(mnt)_2$ (M = Au, Pt)

J. S. Brooks[a*], D. Graf[a], E. S. Choi[a], M. Almeida[b], J.C. Dias[b], R.T. Henriques[b], M. Matos[c]

[a]NHMFL/Physics, Florida State University, Tallahassee, FL 32310, USA

[b]Dept. de Química, Instituto Tecnológico e Nuclear/ CFMCUL , P-2686-953 Sacavém, Portugal

[c]Dept. de Engenharia Química, Instituto Superior de Engenharia de Lisboa, P-1900 Lisboa, Portugal

**Abstract**

The $Per_2M(mnt)_2$ class of organic conductors exhibit a charge density wave (CDW) ground state below about 12 K, which may be suppressed in magnetic fields of order 20 to 30 T. However, for both cases of counter ion $M(mnt)_2$ species studied (M = Au (zero spin) and M = Pt (spin ½) ), new high field ground states evolve for further increases in magnetic field. We report recent investigations where thermopower, Hall effect, high pressure and additional transport measurements have been carried out to explore these new high field phases. © 2001 Elsevier Science. All rights reserved

## 1. Introduction

The α phases of $Per_2M(mnt)_2$ where Per = perylene, mnt = maleonitriledithiolate and M = Au, Pt, (or other transition metals) have been the subject of many studies[1-12] due to the nature of conducting and magnetic instabilities that may arise in the two chains (perylene or $M(mnt)_2$). Here a charge density wave (CDW) instability arises in the perylene chains; when M has a localized magnetic moment( e.g. Pt), a spin Peierls transition may occur. Previous work to 18 T has shown that the CDW transition temperature was decreasing according to mean field predictions[13]. More recently, Graf and co-workers [14, 15] have shown that the low temperature, high resistance CDW phases in $Per_2M(mnt)_2$ (for M = Au, Pt) are suppressed in high magnetic fields of order of the Pauli field $B_P$. A sketch of the magnetic field dependent phase diagrams based on Refs. [14, 15] and the present work is given in Fig. 1.

For fields above $B_P$, the resistance rises again for both the Pt and Au compounds (except for B//b for M = Au). We have described this second, high resistance phase as a field induced charge density wave (FICDW) phase, based on comparisons with current theoretical work in this area[16, 17]. At even higher fields, the FICDW is re-entrant to a second low resistance state for M= Pt. An intervening low resistance state in the vicinity of $B_P$ appears between the CDW and FICDW phases, which for some samples and field orientations appears to be activated. However, in the case of B//c for M=Pt, metallic behavior in the temperature dependent resistance has been was observed below 1.8 K in the vicinity of $B_P$.

In reference to Fig. 1, we emphasize that the phase boundaries depicted are based on both temperature dependent resistance measurements at constant field, and magnetoresistance (MR) measurements at constant temperatures. Hence implicit in the MR data is the notion that as the field destroys or stabilizes a field dependent gap $\Delta(B)$, the MR at constant T will fall or rise with the gap, and from BCS, we assume that $T_c(B) \sim \Delta(B)$. Hence the MR signal vs. B on a log scale will resemble $T_c(B)$ vs. B to a rough approximation. Further extensive R vs. T for increments in B will be needed to unambiguously determine the shape of $T_c(B)$.

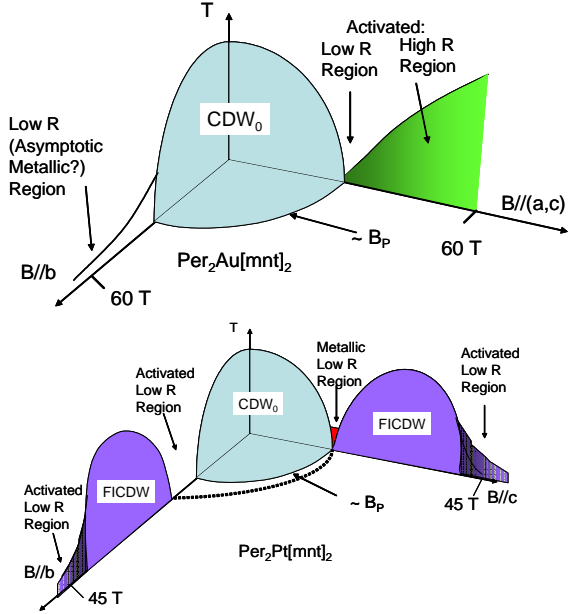

Figure 1. High field behavior of the CDW ground state in $(Per)_2Au(mnt)_2$ and $(Per)_2Pt(mnt)_2$ for $B//c$ and $B//b$ based on Refs. [14, 15] and the present work (see text for discussion).

The purpose of this report is to summarize our results to date on these systems, to present some more recent, albeit, preliminary data, and to put our findings in the context of current theoretical work.

**2. Theoretical background**

Fujita et al. have shown that in the absence of interchain bandwidth, where no orbital effects enter, a commensurate CDW can undergo a transition to an incommensurate ICDW in the presence of a high magnetic field[18]. Here a soliton-like structure can arise in the lattice and charge density, and the ICDW gap will generally decrease with increasing magnetic field. An increase in the magnetization at the CDW-ICDW transition is also predicted[18]. Zanchi, Bjelis, and Montambaux [16] treated a CDW ground state with a highly anisotropic Hubbard model where both spin density wave (SDW) and CDW interactions were included. Predictions representative of their calculations are shown in Fig. 2. An essential feature of the model was the prediction that the ambient ground state $CDW_0$ would evolve into a high field $CDW_X$ state with increasing magnetic field. Moreover, they predicted that as the interchain bandwidth increased from a perfectly nested to a more imperfectly nested state, dramatic differences would arise in the $T_c(B)$ phase diagram. This includes a cascade of transitions when the imperfect nesting parameter is near unity. Indeed, a remarkable comparison between this theory, including the pressure dependence, has been made with the behaviour of the CDW ground state of $\alpha$-(BEDT-TTF)$_2$KHg(SCN)$_4$[19-23]. Most recently, Lebed has treated the theory of a field induced CDW (FICDW) from a metallic, Q1D ground state where, for instance, the CDW ground state is first removed by pressure [17]. A cascade of FICDW transitions are predicted, with a transition temperature lower than for the ambient CDW. Magic angle effects, which change the transition temperature and the frequency of the cascade, are also predicted.

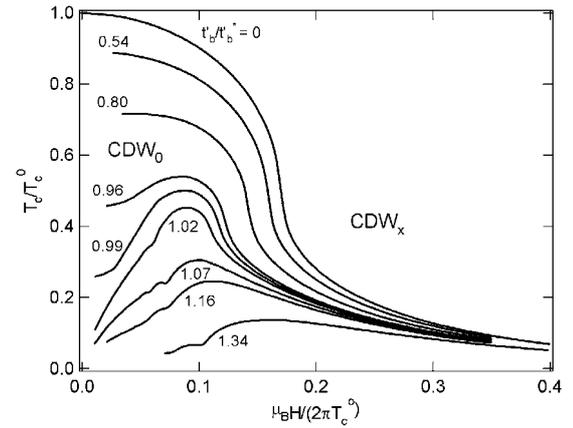

Figure 2. Critical temperature vs. magnetic field for increasing imperfect nesting. (Re-drawn after Ref.[16].)

A general feature of all theoretical work is that in magnetic fields, the nesting vector will be modified by a Zeeman term, and for finite interchain coupling, also an orbital term, with increasing magnetic field.

**3. Experimental results**

The results presented here are made with conventional 4-terminal transport, pressure, and thermopower methods. We note that great care is needed in cooling down the samples, and that 18 hours or more from room temperature to 4.2 K are necessary to insure that the sample will be truly metallic before entering the CDW ground state. Measurements were carried out in 18 T, 33 T, and 45 T superconducting, resistive, and hybrid magnets respectively at the NHMFL and in 60 T at the LANL Pulsed Field Facility.

### 3.1 Per$_2$Au(mnt)$_2$

Referring first to the Per$_2$Au(mnt)$_2$ system, we find that T$_{CDW}$ decreases with increasing field in agreement with both mean field and RPA-type calculations. At, and above the Pauli field B$_P$, the resistance is still thermally activated, although there are dramatic differences in the activation energy with field orientation[14] above B$_P$. Recent pulsed field measurements to 60 T are shown in Fig. 3, which suggest that for B//b (i.e. the field is along the chain direction), the ground state may be metallic in the high field limit, and further experiments are planed to check this possibility. However, when there is a component of the magnetic field perpendicular to the chain direction, a second, high resistance state appears at higher fields, which even at 60 T is still increasing. As noted in section 3.3 below, high pressure shifts this behaviour to lower fields. The M = Au system gave us the first indication that orbital effects may be present in these compounds, even though they are expected to be nearly perfectly one dimensional (see discussion of band structure in Section 4).

We may summarize our understanding of Per$_2$Au(mnt)$_2$ at this stage by noting that our data are consistent with theoretical work predicting ICDW and/or CDW$_x$ states above B$_P$. However, the data deviate from theoretical expectations in the sense that a second high field, high resistance state appears with orientation dependent activation energy. Further temperature dependent work in the 40 to 60 T or greater range is crucial to fully understand this very high field behaviour.

### 3.2. (Per)$_2$Pt(mnt)$_2$

The suppression of the CDW$_0$ ground state in Per$_2$Pt(mnt)$_2$ in high magnetic fields is similar to that in the M=Au compound, but several very important differences appear above B$_P$. The most pronounced difference is that the second, high resistance state (FICDW) that appears above B$_P$ does not continue to increase, but is re-entrant to a lower resistance state in the high field limit. Moreover, unlike the case for M = Au, the FICDW appears for all field orientations, including B//b. Arrhenius analysis from the data of Ref. [15] of the temperature dependent resistance in the FICDW, shown here in Fig. 4, yields activation energies of 12 K, 30 K, and 15 K for fields at B$_P$, at the center of the FICDW, and in the high field re-entrant limit, respectively. Hence the magnetic field induces changes in the FICDW gap that are maximum in the center of the FICDW phase.

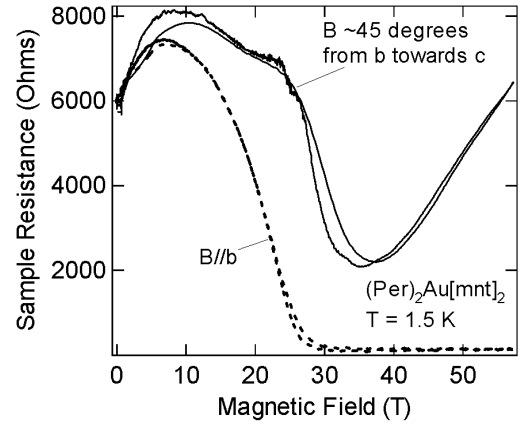

Figure 3. Pulsed field measurements on Per$_2$Au(mnt)$_2$ for two orientations at 1.5 K.

A second phenomena that has only been observed in the M=Pt material, and only for B//c, is that the resistance in the vicinity of B$_P$ ~ 23.5 T drops to a very low value, with a metallic temperature dependence in the low temperature limit, as shown in Fig. 4, again based on the data of Ref. [15]. A subsequent measurement to explore this effect, shown in Fig. 5, has been carried out on another sample. Again, we find a region between the CDW and FICDW phases where metallic behavior can occur. This has only been observed for B//c, i.e. for the field perpendicular to the most conducting plane. Another feature evident in the Per$_2$Pt(mnt)$_2$ compound is the appearance of cascade-like structure in the magnetoresistance, which is most evident in the FICDW phase. We have also seen evidence for this structure in other types of measurements. In Fig. 6 a simultaneous study of two samples, one by magnetization, and the other by electrical transport, are measured as a function of magnetic field for different field orientations. Corresponding structure is seen in both cases. An important issue is that the magnetization is related to the free energy, and hence the structure associated with the FICDW is a bulk thermodynamic phenomena. In Fig. 7, preliminary data where both the resistance and the thermopower of a sample of $Per_2Pt(mnt)_2$ were studied. Here again we find some correspondence between the structure in the two measurements. Fig. 8 shows preliminary data on the Hall effect in $Per_2Pt(mnt)_2$. In a number of Hall measurements, we find an oscillatory structure in the FICDW phase, but its form varies from sample to sample.

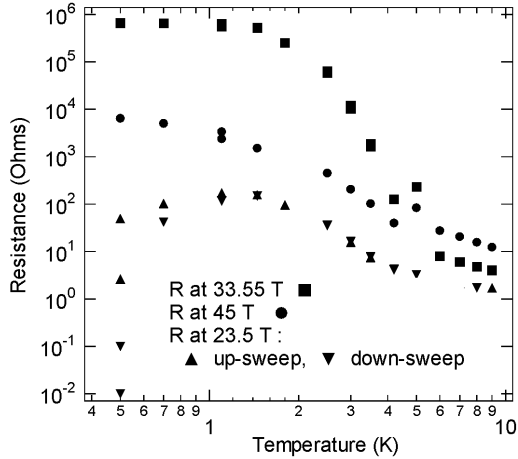

Figure 4. Temperature dependent resistance for $Per_2Pt(mnt)_2$ in the FICDW region. Arrhenius analysis was performed in the 2 to 4 K range of the data. At 23.5 T, the sample becomes metallic below 1.8 K. The scatter in the 0.5 K data for 23.5 T is from measurements from different sweeps through this region. (See text for discussion.)

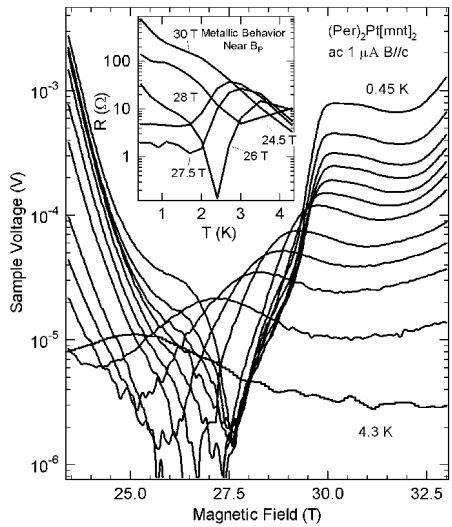

Figure 5. Field and temperature dependence of the resistance of $(Per)_2Pt(mnt)_2$ for B//c in the vicinity of $B_P$ for a second sample.. In the range 24 <B<30 T the resistance shows metallic behavior over a range of temperature and field. dc I-V curves in the vicinity of $B_P$ show Ohmic behavior.

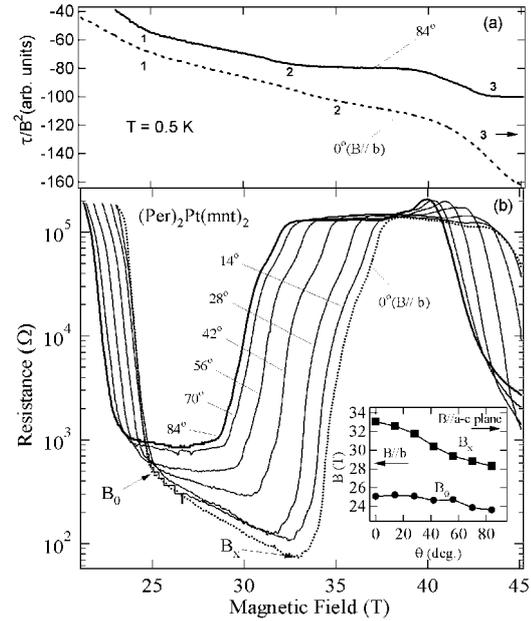

Figure 6. Angular dependence of the FICDW phase in $(Per)_2Pt(mnt)_2$ from B//b ($\theta = 0°$) to B//a-c ($\theta \rightarrow 90°$) at 0.5 K for two different samples. a) Magnetization of the first sample for B//b ($\theta = 0°$) and B//a-c showing the three peak structures. Peak 3 is at too high a field to be observed for B//b ($\theta = 0°$). b) Magnetoresistance of the second sample for the full range of field orientations studied. Note that the B//b ($\theta = 0°$) position was found by continuously rotating the sample holder in field to find the maximum field for $B_x$. Inset: polar angle dependence of the $B_0$ and $B_x$ transitions.

We summarize our understanding of $Per_2Pt(mnt)_2$ by again noting that the suppression of the $CDW_0$ state with magnetic field follows theoretical expectations, but a second, re-entrant FICDW state appears, which is moderately dependent on field direction. Some evidence for orbital-related behaviour appears in the various measurements. The intervening metallic phase near $B_P$ between the $CDW_0$ and FICDW phases for B//c has been reproduced in two samples. Further work is needed to map out its behaviour, and due to the vanishingly small resistance, even superconductivity cannot be ruled out at this time. It is important to realize that the Pt ion carries a spin ½, and NMR measurements[24] show the onset of a spin-Peierls transition at around 50 K. Although high field NMR studies are planned, we do not yet know the field dependence of the spin-Peierls Pt(mnt) system, nor how this will affect the CDW in the Perylene chains.

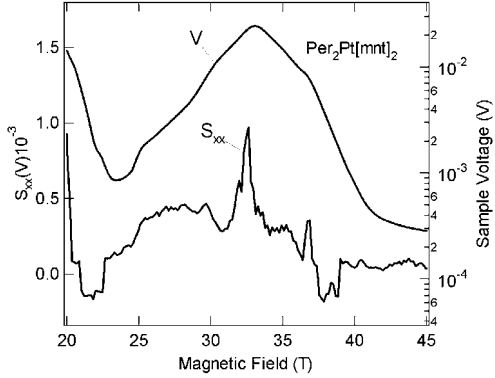

Figure 7. Simultaneous thermopower and resistance measurements of (Per)$_2$Pt(mnt)$_2$ in the region of the FICDW phase.

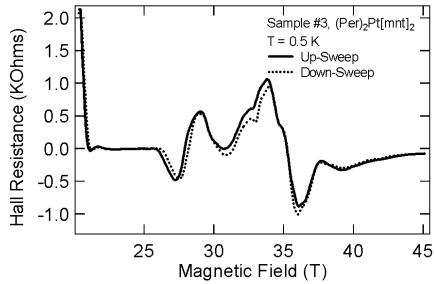

Figure 8. Hall effect in the FICDW state based on subtraction of field forward and field reversed signals (B//c) by sample rotation through 180 degrees for two field sweeps up and two field sweeps down. Finite Hall signals in the FICDW state have been observed in a total of four samples in preliminary studies, but it should be noted that the details vary from sample to sample.

### 3.3. Pressure dependence in (Per)$_2$M(mnt)$_2$

An important aspect of the theoretical work where both spin and orbital effects are considered in the FICDW formation is the variation of the imperfect nesting parameter t'/t'*. This can be accomplished by the application of pressure, which can serve to reduce $T_{CDW}$. For t'/t'* ~ 1, this can lead to the appearance of cascade-like structure in the field dependence of $T_{CDW}$. As discussed above, there has been a remarkable correspondence between the theory and experiment, including high pressure studies, in the case of the system α-(BEDT-TTF)$_2$KHg(SCN)$_4$[19-23]. We are presently making similar high pressure studies on the Per$_2$M(mnt)$_2$ system, and find a similar correspondence with theoretical predictions. Although we have not yet been able to remove the CDW state completely, we have reduced the energy gap and zero field resistance to very low values. The magnetic field dependence of the resistance for both compounds is shown in Fig. 9, where we find that the field is inducing some kind of cascade structure in both compounds.

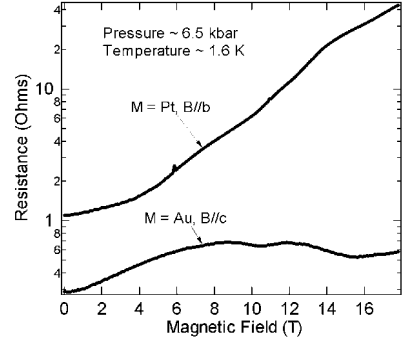

Figure 9. Magnetoresistance of Per$_2$Pt(mnt)$_2$ and Per$_2$Au(mnt)$_2$ at 6.5 kbar at low temperatures.

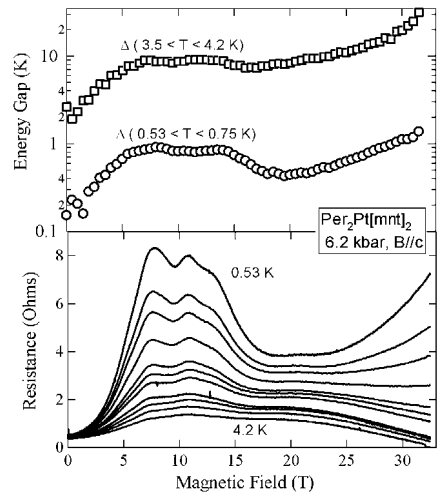

Figure 10. Lower panel: Magnetoresistance of Per$_2$Pt(mnt)$_2$ at 9.5 kbar vs. temperature for B//c. Upper panel: energy gaps at low and high temperatures derived from Arrhenius analysis.

We have also investigated the pressure dependence at higher fields. For M = Au, we find that $B_P$ is reduced to about 14 T at 5 kbar, and that the second high field state again begins above $B_P$ as in Fig. 3 above, but at correspondingly lower fields. For M = Pt, we have made more extensive pressure studies, some of which are shown in Fig. 10. Consistent with theory is the observation of the magnetic field inducing a cascade structure in a CDW system suppressed by high pressure. This comparison can be made by a qualitative comparison of Fig. 2 and Fig. 10. From the temperature dependence, we see that the activation energy is also a function of the magnetic field. At present, the cascade structure does not appear to be periodic in inverse field, and from angular dependent studies, is only approximately dependent on the component of magnetic field perpendicular to the a-b plane. We also note that unlike the theory (Fig. 2), we still observe a second increase into a second high field state in both compounds, even under pressure.

## 4. Discussion

Although $Per_2M(mnt)_2$ is thought to be a nearly perfect one-dimensional conductor along the perylene chains, our angular dependent magnetotransport data indicate the possible influence of orbital effects. To more carefully examine this issue, the electronic structure of the $Per_2M(mnt)_2$ system has recently been re-examined by Canadell et al., [25]. Computations with single-$\zeta$ integrals showed a large ratio between the intrachain (~150 meV for the b-axis) to interchain (~ 2 meV for the a-axis; ~0 meV for the c-axis) bandwidths. Although the double-$\zeta$ bandwidths were uniformly about a factor of two larger, the ratios were very similar. From the band structure, if there is a "most conducting plane", it will be the a-b plane, where we note the a-axis bandwidth is very small, more than an order of magnitude less than those, in for instance, the Bechgaard salts[26] which show field induced spin density wave (FISDW) behavior and clear $1/\cos(\theta)$ tilted field dependence of the field induced structure. It was noted however[25], that there are four inequivalent perylene chains in the unit cell, which in the absence of hybridization, may lead to multiple, slightly warped Fermi surface sheets. Although the energy scale of this topology is of order meV, the thermal and magnetic phenomena described herein are also on this scale. Hence it is possible that the deviations from perfect one-dimensional electronic structure lead to the appearance of orbital effects in the field dependence of the CDW ground state. It is also possible that if there are several bands at the Fermi level, the Zeeman splitting of the degenerate bands may first begin to destroy the CDW nesting condition. However, for higher fields, there may be a coincidence where different spin split bands overlap, and the nesting condition improves. This could lead to the second (FICDW) phase at higher fields. Eventually even this nesting condition would degrade, leading to the re-entrant behaviour, as seen in the M=Pt system.

*Note in proof:* Mitsu et al. [27] have recently reported a non-CDW, non-metallic state for M = Pt above 5 kbar at low temperatures (without magnetic field).

**Acknowledgements**
Supported by NSF-DMR 02-03532 and a NSF GK-12 Fellowship (DG). NHMFL is supported by the NSF and the State of Florida. Work in Portugal was supported by FCT under contract POCT/FAT/39115/2001.